\documentclass[conference]{IEEEtran}
\IEEEoverridecommandlockouts
\usepackage{cite}
\usepackage{amsmath,amssymb,amsfonts}
\usepackage{mathtools}
\usepackage{algorithmic}
\usepackage{wrapfig}
\usepackage{textcomp}
\usepackage{xcolor}
\usepackage{url} 
\usepackage{soul,color}
\usepackage{adjustbox}
\usepackage{comment}
\usepackage[hidelinks]{hyperref}
\usepackage{orcidlink}
\usepackage{multirow} 
\usepackage{enumitem}
\usepackage{svg}
\usepackage{subcaption}
\usepackage{balance}

\usepackage[capitalize]{cleveref}
\crefname{section}{Sec.}{Secs.}
\Crefname{section}{Section}{Sections}
\Crefname{table}{Table}{Tables}
\crefname{table}{Tab.}{Tabs.}

\def \etal {\emph{et al.}}

\begin{document}

\title{Feasibility Assessment of Remote Driving via Latency Analysis of ITS-G5 and Cellular Networks in the MASA Living Lab}

\author{\IEEEauthorblockN{
Gaetano Orazio Cauchi$^1$,
Antonio Solida$^1$,
Salvatore Iandolo$^1$,
Marco Savarese$^1$,
Martin Klapez$^1$,\\
Enrico Rossini$^2$,
Marcello Pietri$^2$,
Marco Picone$^2$,
Marco Mamei$^2$,
Maurizio Casoni$^1$,
Carlo Augusto Grazia$^1$}
\IEEEauthorblockA{
\textit{$^1$Department of Engineering ``Enzo Ferrari'', $^2$Department of Science and Methods for Engineering,}\\
\textit{University of Modena and Reggio Emilia}\\
\{name.surname\}@unimore.it
}
}

\maketitle

\begin{abstract}
Remote driving has gained increasing attention as a key enabler for connected and automated vehicles. Yet its practical deployment hinges on wireless networks' ability to guarantee low, predictable latency. In this paper, we present an extensive latency analysis of ITS-G5 and cellular (5G) technologies within the Modena Automotive Smart Area (MASA), a real-world, city-scale testbed equipped with a distributed intelligent transportation infrastructure. By conducting controlled experiments under varying network loads and traffic conditions, we measure network and end-to-end latency components relevant to remote driving, in which the uplink consists of a continuous video stream transmitted from the vehicle to the remote operator, and the downlink conveys control commands back to the car. Measurements conducted under diverse conditions reveal how latency and variability differ across the two technologies and how infrastructure coverage impacts video-stream transmission performance. Based on the observed latency distributions and reliability metrics, we assess the practical feasibility and safety margins of remote driving in mixed network environments. The results provide actionable insights for future teleoperation deployments and motivate hybrid communication strategies that combine the strengths of ITS-G5 and cellular networks.
\end{abstract}


\section{Introduction}
In recent years, the evolution of Intelligent Transportation Systems (ITS) has been strongly influenced by regulatory initiatives such as the European Directive 2010/40/EU~\cite{eu_directive}, which aims to improve road safety, traffic efficiency, and overall mobility. A key enabler of ITS is the exchange of information among vehicles, infrastructure, and other road users. In this context, Vehicle-to-Infrastructure (V2I) communication has opened new perspectives, including the possibility of remotely controlling vehicles through command transmission from geographically distant operators, thus enabling remote driving applications.

Despite its potential, remote driving imposes stringent requirements on communication latency and reliability. A remote operator must perceive and respond to dynamic traffic conditions with minimal delay, which requires ultra-low-latency bidirectional communication: a downlink for vehicle control commands and a reliable uplink for video to ensure real-time situational awareness. In particular, video latency is critical, as excessive delays can impair the driver’s ability to judge distances and react accurately to hazards. For this reason, remote driving systems typically target end-to-end latencies of $100–150ms$~\cite{zhang2020toward}.

This paper presents an experimental evaluation of remote driving over an ITS-G5 (IEEE 802.11p) infrastructure deployed in the Modena Smart Automotive Area (MASA) living lab, a real-world urban testbed featuring a dense network of ETSI-compliant Roadside Units (RSUs) interconnected via a fiber-optic backbone. The experimental setup focuses on uplink video streaming from a vehicle equipped with a dedicated communication platform supporting both IEEE 802.11p and cellular connectivity. Real-time video is transmitted using MPEG Transport Stream over UDP multicast with M-JPEG encoding, allowing multiple RSUs to receive the stream concurrently without increasing wireless load. Cellular 5G New Radio connectivity is used as a fallback to ensure transmission continuity outside RSU coverage.

Unlike prior studies that rely on theoretical latency models, this work reports empirical Glass-to-Glass latency measurements from a live hybrid V2X network. The data show that ITS‑G5 achieves lower latency where RSU coverage is available, but cellular fallback remains necessary to maintain service continuity in complex urban layouts. These results confirm that hybrid V2X architectures are practically viable for remote driving deployment.

 \begin{figure*}
   \centering
    \includegraphics[width=0.88\textwidth, trim = {0cm 0cm 0cm 0cm}]{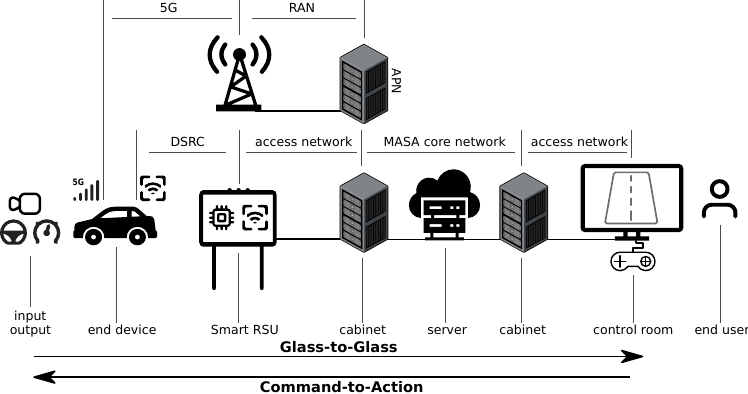}
     \caption{MASA: real living-lab network.}
     \label{fig:masa}
 \end{figure*}

\section{Related Works}
\label{sec:related}
Remote driving lies between human-driven vehicles and full autonomy, but it faces key challenges, including communication latency, limited situational awareness, and insufficient feedback to remote operators~\cite{zhao2024remotesurvey}. Human-centric aspects, including operator experience, control interfaces, and usability, are also critical to system effectiveness~\cite{li2024remotefailsafe}.

Several studies address the technical feasibility of remote driving. Kang~\etal~\cite{kang2018augmenting} analyze latency, safety, and the integration of teleoperation with autonomous functions. Yu~\etal~\cite{yu2022remotecontrol} study control strategies under network constraints, while Heryana~\etal~\cite{heryana2023realtimevideo} demonstrate low-latency video streaming using UDP. Together, these works highlight the importance of timely video delivery for maintaining control and situational awareness.

Most experimental systems rely on Wi-Fi or cellular networks (4G/5G), which often suffer from variable latency. Liu~\etal~\cite{liu2017investigating} show that LTE delay variability can degrade teleoperation performance. Prior work indicates that teleoperation is more tolerant to small, stable delays than to highly variable latency~\cite{chen2007hpi, liu2017investigating, davis2010timelag}, motivating the use of vehicular communication technologies with more predictable latency.

Low-latency video transmission has been explored in vehicular networks. Marai~\etal~\cite{marai2020smooth} emphasize its importance for remote driving and platooning, while Huang~\etal~\cite{huang2024supporting} analyze bandwidth requirements for 720p video at $30~fps$. Kamtam~\etal~\cite{kamtam2024networklat} provide a comprehensive review of the effects of network latency on vehicle teleoperation and related mitigation strategies.

Experimental studies such as~\cite{pereira2018v2v, win2020road, kamtam2024networklat} evaluate real-time video streaming over IEEE~802.11p-based V2V communications and identify challenges in reliability, throughput, and latency. Brahim~\etal~\cite{brahim2017qos} further define packet loss, jitter, and latency as key QoS metrics for vehicular video streaming.

Finally, Noguchi~\etal~\cite{noguchi2011location} propose IPv6 multicast over GeoNetworking for efficient geographic data dissemination, an approach partially adopted in this work to support real-time video distribution to multiple RSUs. This study builds on preliminary measurements of the MASA\footnote{https://www.automotivesmartarea.it/}ITS infrastructure by Grazia~\etal~\cite{grazia2024preliminary} and on the real-world evaluation of low-latency vehicle–infrastructure communication for remote driving by Solida~\etal~\cite{solida2025remote}.

\section{Methodology and Experimental Setup}
\label{sec:methandexp}
\subsection{The MASA Living Lab}

The Modena Automotive Smart Area (MASA) is an open-air urban living laboratory designed to support the experimentation, validation, and deployment of connected and cooperative mobility technologies under realistic traffic conditions. The testbed spans portions of Modena’s urban environment and university campus, combining public roads and controlled areas with heterogeneous layouts to enable comprehensive evaluation of Intelligent Transportation Systems (ITS) and Vehicle-to-Everything (V2X) applications. The central part of~\Cref{fig:masa} depicts the MASA core network.

MASA features a dense deployment of ETSI-compliant Roadside Units (RSUs), strategically positioned to provide continuous V2X coverage across the test area. The RSUs are interconnected via a high-capacity optical fiber backbone that forms the MASA core network, ensuring low-latency, high-reliability connectivity among roadside infrastructure, edge computing nodes, and backend services. This architecture enables hybrid communication paths, where time-critical data are exchanged over short-range wireless links, while computation-intensive tasks are offloaded to edge or centralized resources.

The RSUs deployed within MASA are industrial-grade units manufactured by Movyon Electronics and are fully compliant with European ITS-G5 (IEEE 802.11p) standards and the C-V2X specifications defined in 3GPP Releases 14 and 15. Each RSU integrates a dedicated ITS-G5 software stack and operates in the regulated 5.9 GHz ITS band, providing reliable V2X communications with a reception sensitivity of -97 dBm. Accurate georeferencing is ensured through multi-constellation GNSS support (GPS, Galileo, GLONASS, and BeiDou), enabling precise positioning for cooperative awareness and infrastructure-assisted applications. The devices are CE RED certified and housed in IP67 enclosures for outdoor operation.

A distinguishing feature of MASA is the tight integration between the roadside access network and Multi-access Edge Computing (MEC) resources. Edge nodes connected to the MASA backbone host latency-sensitive services, including packet deduplication, real-time video processing, cooperative perception, and data aggregation. Furthermore, the MASA core network supports both unicast and multicast data distribution, enabling efficient dissemination of high-bandwidth streams—such as video feeds—for applications including remote driving, traffic monitoring, and cooperative perception.

\begin{figure}
    \centering
    \begin{subfigure}[t]{.85\columnwidth}
    \vspace{0pt}
    \centering
    \includegraphics[width=\textwidth,
            height=0.55\textheight,
            keepaspectratio]{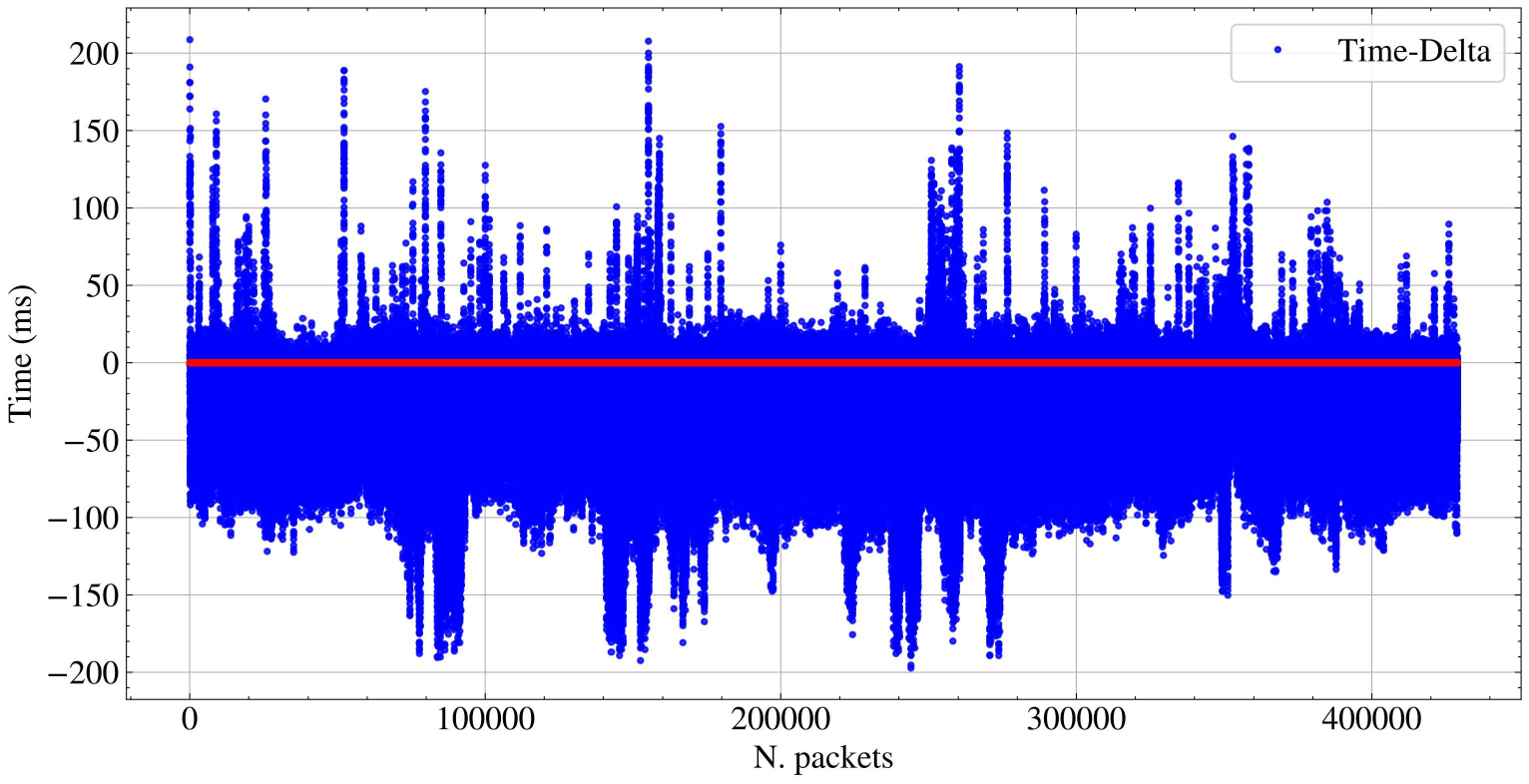}
    \caption{}
    \label{fig:scatter_time_deltas}
\end{subfigure}
\hfill
\begin{subfigure}[t]{.13\columnwidth}
\vspace{0pt}
    \centering
    \includegraphics[width=\textwidth,
            height=0.4\textheight,
            keepaspectratio]{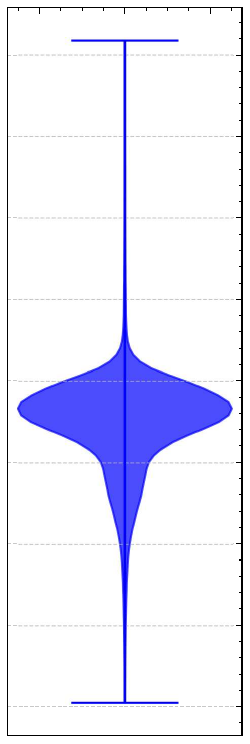}
    \caption{}
    \label{fig:violin_time_deltas}
\end{subfigure}
    \caption{Delta time comparison between DSRC and 5G: sequence plot (a) and violin distribution (b).}
    \label{fig:plots_time_deltas}
\end{figure}

\subsection{Technologies | Hardware setup}
\label{subsec:techhardware}

The remote driving experiments conducted within the Modena Automotive Smart Area (MASA) rely on a heterogeneous hardware platform that supports real-time video uplink and reliable bidirectional connectivity between vehicular and infrastructure networks. The core computing unit deployed on the vehicle is the Vault V1410 from Protectli (hereafter Vault), a compact x86-based edge device suitable for multi-interface V2X and teleoperation scenarios.

For short-range vehicular communications, Vault integrates a MikroTik R11e-5HnD wireless card, an IEEE 802.11a/n adapter based on the Qualcomm Atheros AR9580 chipset, equipped with a passive heat sink for continuous operation. While initially designed for Wi-Fi communications, the AR9580 chipset natively supports operation in the upper 5 GHz spectrum. Support for IEEE 802.11p/ITS-G5 communications in the 5.9 GHz band is enabled via software-level extensions, including custom patches to the Linux ath9k driver and regulatory domain configuration. This allows the device to operate on ITS channels and participate in V2X communication experiments within MASA.

In addition to IEEE 802.11p connectivity, Vault is equipped with a Quectel RM520N-GL cellular modem that supports 5G NR and LTE natively. In the current deployment, the modem operates on 5G NR connectivity, as this is the cellular service currently provided by the Municipality of Modena across the MASA area. The modem uses a dedicated SIM card configured with a municipal APN, ensuring seamless integration with the MASA networking infrastructure. This configuration provides continuous connectivity between the vehicle, MASA Roadside Units (RSUs), edge nodes such as HAura, and the central Multi-access Edge Computing (MEC) platform.

Vehicle localization during the experiments is supported by an external USB-based GNSS receiver, directly connected to the Vault. The GNSS module provides real-time positioning data used to georeference experimental measurements and to correlate network performance metrics with the vehicle’s spatial trajectory throughout the MASA area.

Video acquisition is performed using a Raspberry Pi Camera Module 3, interfaced to a dedicated capture subsystem connected to the Vault. The camera integrates a 12-megapixel Sony IMX708 image sensor, which offers high spatial resolution, and an autofocus mechanism that enables dynamic focus adjustment, ensuring stable image quality across varying distances and lighting conditions.

\subsection{Technologies | Software setup}
\label{subsec:softwaresetup}
The remote driving experiments are structured around a dual-connectivity communication model designed to ensure continuous video delivery and to enable a comparative evaluation of short-range IEEE 802.11p and wide-area cellular technologies within the MASA environment.

The uplink channel is primarily used for real-time video streaming from the vehicle to the MASA infrastructure. Video frames captured by the onboard camera are transmitted via MPEG Transport Stream (MPEG-TS) over UDP multicast, with frames encoded in Motion JPEG (M-JPEG). This encoding choice balances computational complexity and visual quality, enabling real-time operation while preserving sufficient image fidelity for remote driving tasks.

Multicast transmission is deliberately adopted at the wireless level to improve scalability and efficiency. By using multicast, all RSUs within radio range can simultaneously receive the same video stream without introducing additional load on the IEEE 802.11p channel. This approach is particularly advantageous in dense RSU deployments, where multiple infrastructure nodes may be in the vehicle's line of sight.

At the networking level, multicast video packets are encapsulated as IPv6 multicast datagrams over GeoNetworking, in compliance with ETSI specifications \cite{etsiTS102636-6-1}. Each RSU that receives the multicast stream forwards the packets over the MASA fiber-optic backbone toward the processing and visualization infrastructure, ensuring low-latency, high-reliability transport within the core network.

\begin{figure}[t]
        \centering
        \includegraphics[width=\linewidth]{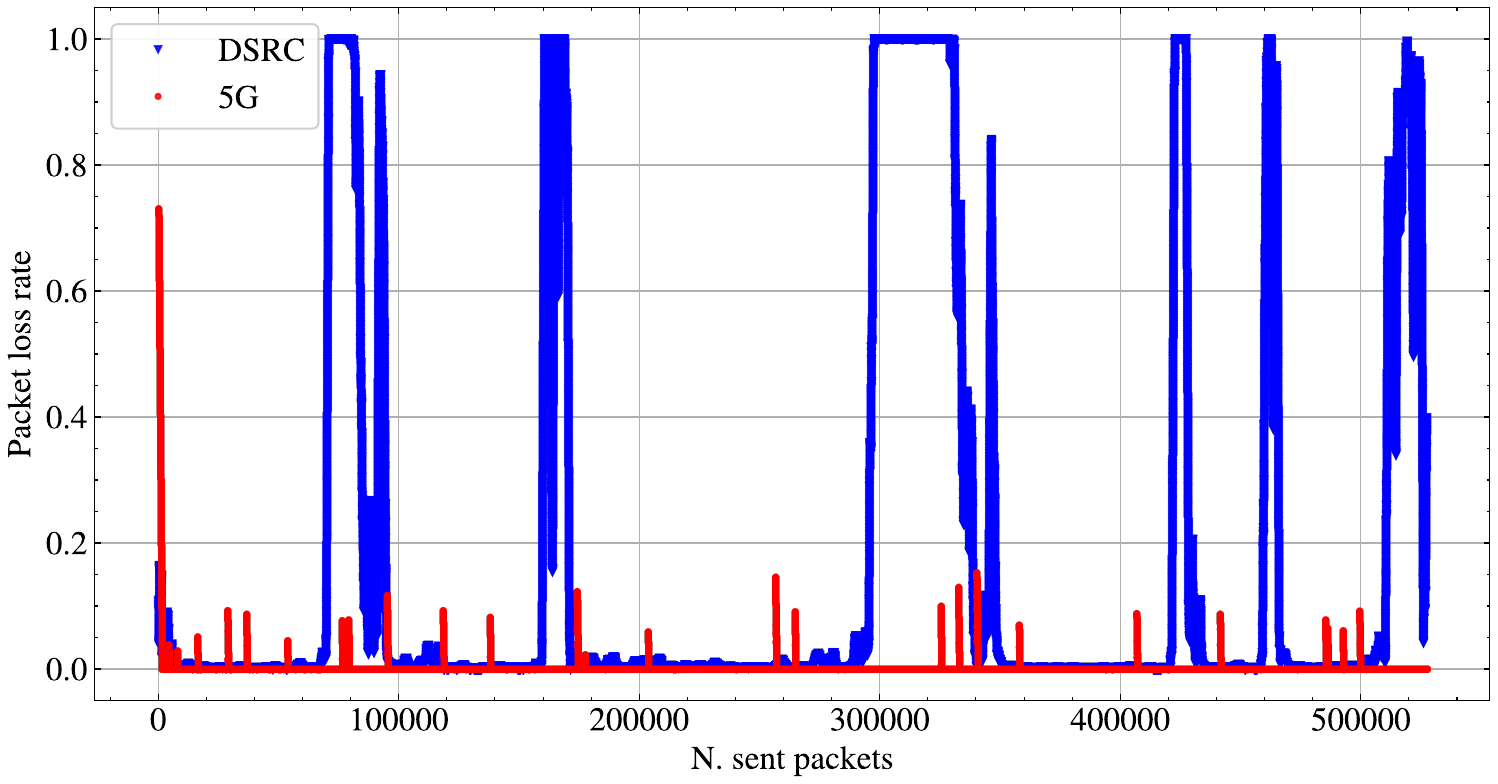}
        \caption{Loss-Packet rate comparison between DSRC and 5G.}
        \label{fig:loss-rate}
\end{figure}

The experimental campaign includes vehicle trajectories spanning different areas of MASA, covering both zones with IEEE 802.11p RSU coverage and sections without DSRC infrastructure. In the latter case, uplink video packets are delivered directly to the MEC node through the 5G cellular link, leveraging the Vault platform’s cellular modem.

In summary, video data are transmitted simultaneously through:
\begin{enumerate}
    \item IPv6 multicast packets over GeoNetworking, received by nearby RSUs when DSRC coverage is available;
    \item UDP unicast packets over the 5G network, directly addressed to the MEC node.
\end{enumerate}

The overall data transmission flow, including both DSRC-based multicast and 5G unicast paths toward the MEC, is illustrated in~\Cref{fig:masa}.

Both communication paths terminate within the same logical MASA network, enabled by a municipal SIM card and APN configuration. This guarantees full IP reachability between the vehicle, RSUs, edge nodes, and the MEC, regardless of the access technology in use. The dual-path strategy ensures service continuity while allowing the performance of DSRC and cellular links to be evaluated under identical operational conditions.

When the video stream is received via the DSRC path, multiple RSUs may forward identical packets to the MEC node due to overlapping coverage areas. To mitigate redundant traffic and prevent unnecessary processing, the MEC incorporates a lightweight packet deduplication mechanism.

The deduplication module operates on a UDP socket at the MEC, where a CRC32 checksum is computed over the payload of each packet forwarded by the RSUs. CRC32 was selected for its minimal computational overhead and sufficient collision resistance for latency-sensitive applications such as remote driving, enabling real-time processing without affecting end-to-end latency.

Packet hashes are stored in a fixed-size circular cache to identify duplicates. Packets whose hashes are not present in the cache are forwarded to the visualization pipeline. At the same time, duplicates are discarded, ensuring that only a single copy of each packet propagates beyond the MEC. This mechanism preserves stream stability and prevents bandwidth waste caused by overlapping RSU coverage. Packets received via the 5G unicast path bypass deduplication, as they follow a separate delivery channel.

At the receiving end, the MEC forwards the video stream to a remote operator workstation via unicast, where it is decoded and displayed in real time using an FFmpeg-based pipeline. The system also supports application-level multicast reception, allowing multiple remote clients to view the same stream concurrently without duplicating upstream traffic.

In parallel with video streaming, the vehicle periodically transmits ETSI-compliant Cooperative Awareness Messages (CAMs) that contain GNSS-derived information, such as position, speed, and heading. These messages are disseminated over the IEEE 802.11p interface and collected by the MASA infrastructure, providing situational awareness and enabling spatial correlation between network performance metrics and vehicle dynamics.

\subsection{Glass-to-Glass Latency}
\label{subsec:g2gmeth}
Although often overlooked, glass-to-glass latency is a critical metric for safety-critical applications such as remote driving. This measure captures the total time required for a video frame to propagate from the camera sensor (the first glass) to the display screen (the second glass). While network latency provides valuable insight into communication delays, it does not fully characterize system performance, as factors such as camera frame rate, encoding and decoding pipelines, and video codecs can significantly contribute to the overall delay.

We adopted a dedicated framework for glass-to-glass latency evaluation to assess our experimental setup. The testbed consists of an LED light source, a photodiode, an ESP32 microcontroller interfacing with both components, the camera described in~\Cref{subsec:techhardware}, and a $120$ Hz display. Video streams are encoded and decoded using the H.264 codec.

Overall, the measurements quantify the time required for a video frame in our system to propagate from camera acquisition to its visualization on the monitor. A detailed description of the experimental methodology and the obtained results is provided in~\Cref{subsec:resg2g}.

\section{Results}
The experimental results presented in this section were obtained using the methodology and dual-connectivity setup detailed in~\Cref{sec:methandexp}. Drive tests were conducted at an average speed of $30~km/h$ over multiple laps within the MASA living lab. The selected route traverses varying RSU-density zones and 5G signal profiles, ensuring a comprehensive evaluation of network transitions. Measurements remained consistent across all runs, with no appreciable differences observed. Raw traces were processed to extract time deltas, packet loss rates, bitrate fluctuations, and RSU coverage statistics. To isolate video processing delays from wireless channel effects, Glass-to-Glass latency was characterized separately in a controlled laboratory environment.

\begin{figure}
  \centering
\includegraphics[width=0.50\textwidth]{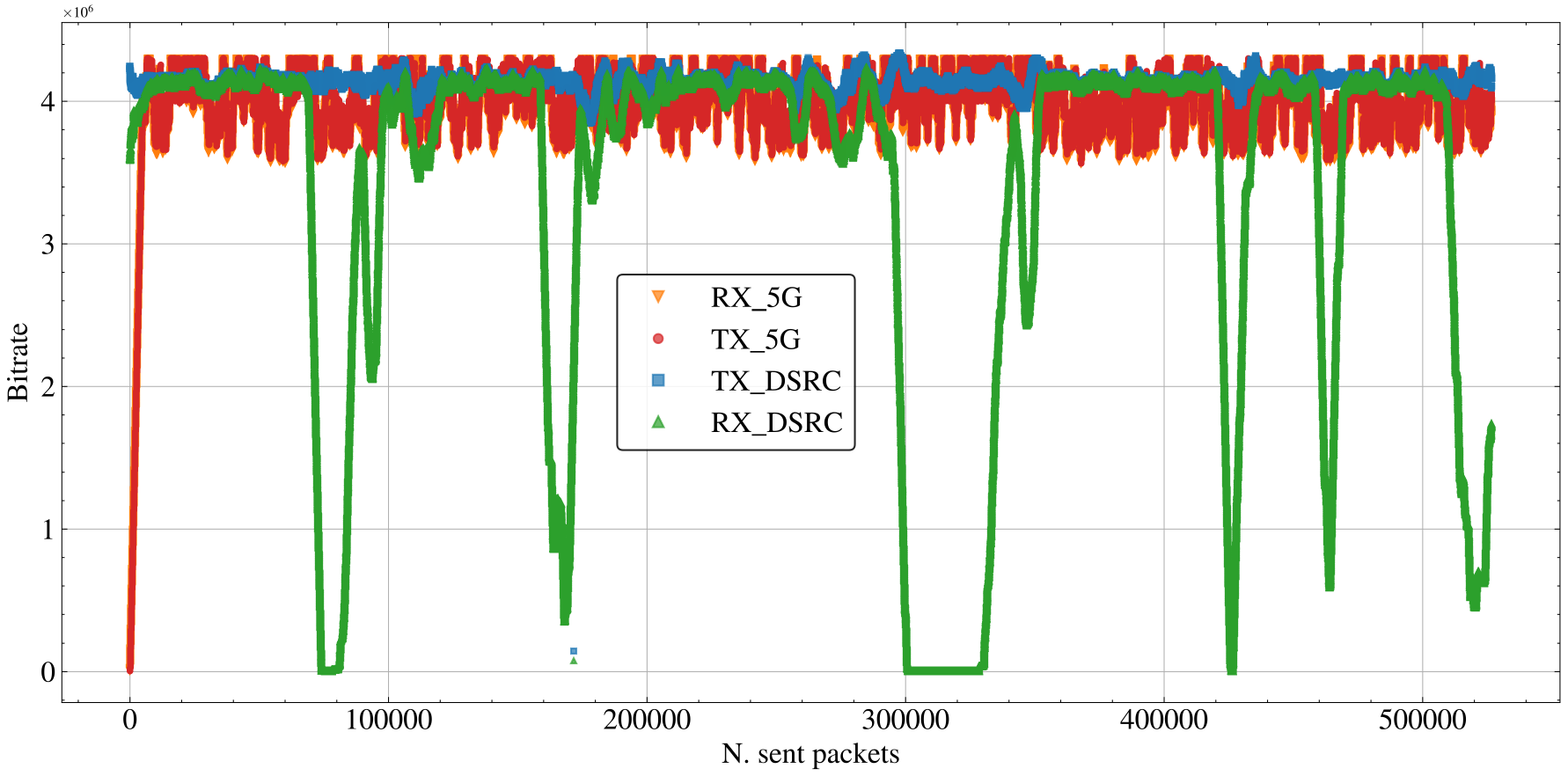}
    \caption{Transmitted vs. Received Bitrate, both DSRC and 5G.}
    \label{fig:bitrate}
\end{figure}

\begin{figure*}[t]
    \centering
    \begin{subfigure}{0.36\textheight}
        \centering
        \includegraphics[width=0.99\linewidth]{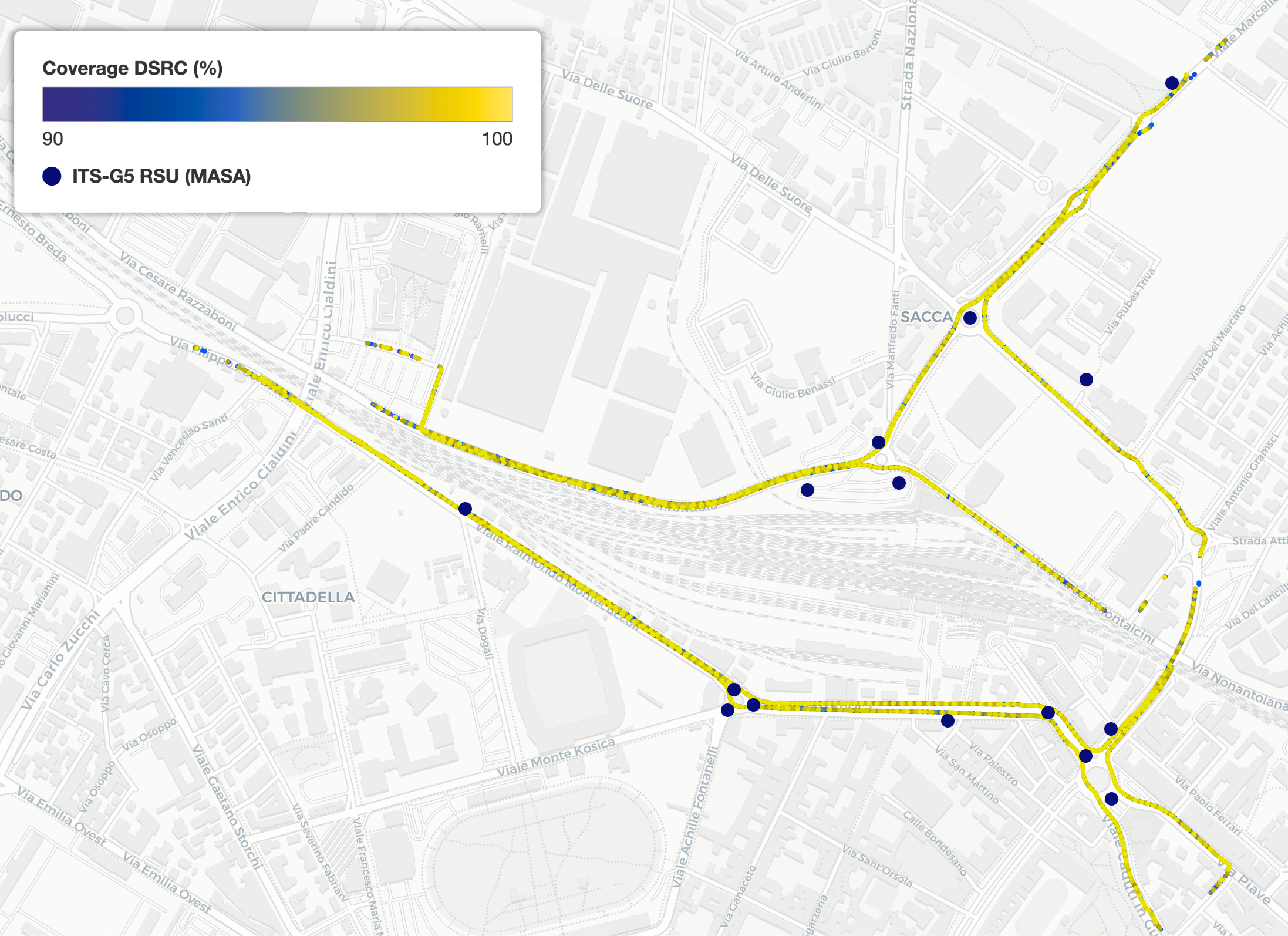}
        \caption{DSRC Coverage map along with RSUs positions.}
        \label{fig:path_dsrc}
    \end{subfigure}
    \hfill
    \begin{subfigure}{0.36\textheight}
        \centering
        \includegraphics[width=0.99\linewidth]{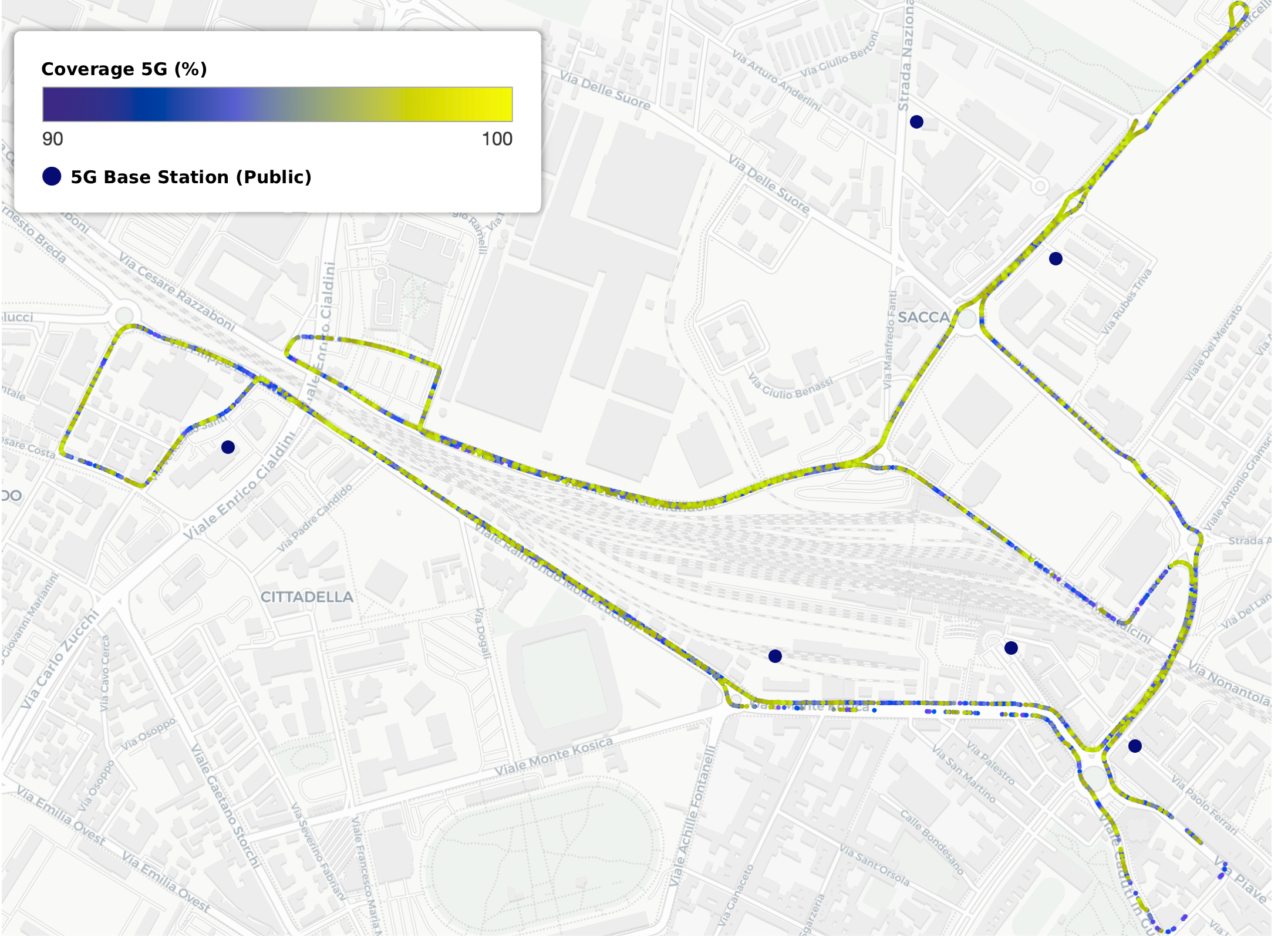}
        \caption{5G Coverage map along with Base Stations positions.}
        \label{fig:path_4g}
    \end{subfigure}
    \caption{Maps of the test scenario.}
    \label{fig:coverage_group}
\end{figure*}
\subsection{Time-Deltas}

The first metric we evaluated to compare cellular networks (5G) and ITS-G5 (DSRC) is the time interval between the packet arrival time using one technology and the same packet's arrival time using the other technology (Time-Deltas). This comparison is possible and meaningful because each packet is transmitted simultaneously by both technologies.

Time deltas are calculated as follows: 

\[
\Delta Time = t_{\text{5G, i}} - t_{\text{DSRC, i}}
\]

Therefore, if $\Delta Time > 0$, the cellular transmission was faster; if $\Delta Time < 0$, DSRC was faster. As shown in~\Cref{fig:plots_time_deltas}, $96.7\%$ of the values are negative, confirming DSRC’s latency advantage within coverage. The positive spikes correspond to coverage transitions or interference events, where 5G provides more stable delivery. While DSRC outperforms 5G in latency, it operates in a shared ITS band that is susceptible to contention. Conversely, 5G cellular spectrum in Italy remains underutilized due to low commercial penetration, resulting in lower channel congestion. Given the current deployment landscape in Italy, both technologies remain directly comparable, enabling a balanced evaluation of their trade-offs for hybrid remote driving systems.

\subsection{Loss-Packet Rate}

The packet loss rate is a metric used to evaluate the reliability of a communication system based on a given technology.

As shown in Figure \ref{fig:loss-rate}, 5G exhibits more consistent performance across the entire scenario considered, in line with its high geographical coverage and the presence of a widespread cellular infrastructure. In contrast, DSRC has more limited coverage, constrained by the presence of RSUs, but in areas where communication is effectively supported, the loss packet rate is low and stable. This behavior indicates that, under guaranteed radio coverage, DSRC can offer high connection reliability, thanks to direct communication and the low latency typical of ad hoc vehicular networks.

The analysis of the loss packet rate, therefore, shows a clear trade-off between coverage and connection quality. While 5G guarantees continuous connectivity at a large scale, DSRC ensures exceptionally reliable communication performance in areas with RSU infrastructure. This result supports the idea of a complementary or hybrid approach between the two technologies to meet the requirements of advanced vehicular applications.

\subsection{Bitrate}
The bitrate analysis aims to characterize the performance of the proposed remote driving communication architecture under real operating conditions, considering both ITS-G5 (DSRC) and cellular (5G) technologies. ~\Cref{fig:bitrate} reports four bitrate profiles: DSRC transmission (TX), DSRC reception (RX), 5G transmission (TX), and 5G reception (RX), computed during continuous video streaming from the vehicle to the MEC.

An additional challenge in bitrate estimation arises from intermittent coverage gaps affecting both DSRC and 5G links. In such conditions, the receiver may experience time intervals with no packets received, which would otherwise lead to inconsistent comparisons between transmitted and received bitrates. To address this issue, the bitrate calculation enforces a one-to-one correspondence between transmitted and received packets for both technologies. Specifically, for each transmitted packet identified by a unique payload–timestamp pair, the receiver checks whether the same payload is observed within a 0–100 ms temporal window. If the packet is received within this interval, it is considered successfully delivered and contributes to the RX bitrate. Otherwise, the packet is classified as either late or lost, and a zero-payload sample is inserted at the receiver side. This method preserves temporal alignment between TX and RX streams and enables a coherent bitrate comparison.

The resulting bitrate profiles show that both DSRC and 5G technologies maintain stable, consistent throughput during active video streaming, with TX and RX bitrates closely matching under coverage conditions. Temporary drops in the DSRC RX bitrate are observed when the vehicle exits the RSUs' coverage area, producing downward spikes that are entirely consistent with the underlying radio conditions. In these intervals, 5G connectivity provides transmission continuity, ensuring uninterrupted video delivery to the MEC.

Overall, the bitrate analysis confirms the effectiveness of the hybrid communication approach: DSRC provides high-performance, low-latency video delivery within RSU coverage, while 5G serves as a complementary fallback, maintaining service continuity in uncovered areas without significant degradation in streaming performance.

\subsection{Coverage comparison}
One of the purposes of these tests was to quantify wireless coverage quality, defined as the percentage of packets correctly received by the MEC node compared to those sent by the vehicle; this metric was then spatially projected onto a map to visualize the actual extent of the service and identify critical issues related to the urban environment. As illustrated in Figure \ref{fig:coverage_group}, the analysis focused on a direct comparison between two technologies used in this work: DSRC (Dedicated Short-Range Communications) and the 5G cellular network.

The methodology involved systematically moving the vehicle within the area of interest along the path indicated on the map. During the journey, the system continuously transmitted a video stream with characteristics comparable to those of the remote driving service. At the same time, CAM messages were used as specified in~\Cref{subsec:softwaresetup} to provide periodic position updates and to synchronize video packets with geographic location.

The coverage behavior of each technology is discussed below.

\begin{itemize}
    \item \textbf{DSRC Coverage} (\Cref{fig:path_dsrc}): Experimental connectivity analysis reveals high spatial reliability along the monitored road segments. The ITS-G5 network delivers continuous coverage, with Packet Delivery Ratio (PDR) values approaching 100\%. Signal stability extends beyond immediate RSU proximity into interconnected sections, indicating that node placement effectively preserves Line-of-Sight (LoS) along primary routes. Local performance degradation occurs only at specific intersections and segments where architectural obstructions introduce Non-Line-of-Sight (NLoS) conditions and multipath fading. These results confirm the corridor-centric nature of ITS-G5: while performance remains excellent along covered routes, achieving widespread urban coverage would require further infrastructure densification to compensate for the inherent propagation limitations of 5.9 GHz in NLoS environments.
    
    \item \textbf{5G coverage} (\Cref{fig:path_4g}): In contrast, 5G coverage spans the entire test route by utilizing existing commercial cellular infrastructure, without requiring dedicated roadside units. This architecture guarantees wide-area connectivity but yields heterogeneous link performance. While most segments sustain Packet Delivery Ratio (PDR) values near 100\%, localized drops to approximately 90\% frequently occur in dense urban canyons and complex intersections. These fluctuations are primarily driven by radio-resource contention in the commercial network and by building-induced shadowing or attenuation. Missing measurement points in the southern route sections do not indicate complete signal loss; rather, they correspond to intervals in which channel quality fell below the data-logging threshold, resulting in intermittent connectivity gaps.
\end{itemize}

The coverage maps clearly reflect this behavioral difference: ITS-G5 exhibits a quasi-binary spatial profile, characterized by abrupt transitions between high-reliability zones and coverage gaps, whereas 5G shows a graded, more continuous signal distribution. These results support a targeted hybrid deployment strategy. For mission-critical applications such as remote driving, a multi-RAT architecture provides the most robust framework: 5G ensures wide-area baseline connectivity, while ITS-G5 can be selectively deployed at critical intersections or high-risk segments where deterministic, ultra-low-latency is required.

\subsection{Glass-to-Glass Latency}
\label{subsec:resg2g}

The glass-to-glass latency is measured following the methodology described in~\Cref{subsec:g2gmeth}. During the experiment, the ESP32 microcontroller activates the LED every two seconds and records a timestamp at each activation. The camera captures the illuminated LED, and the resulting video stream is displayed on the monitor. A photodiode is affixed to the monitor at the location corresponding to the LED’s on-screen position. When the LED becomes visible on the display, the photodiode is triggered and a second timestamp is recorded.

The difference between the two timestamps represents the glass-to-glass latency, encompassing sensor acquisition, video encoding and decoding, transmission, and display refresh delays. These measurements complement the network latency results, providing a comprehensive assessment of the system’s end-to-end performance.

\begin{figure}
    \centering
    \begin{subfigure}[t]{.85\columnwidth}
    \centering
    \includegraphics[width=\textwidth, height=0.55\textheight ,keepaspectratio]{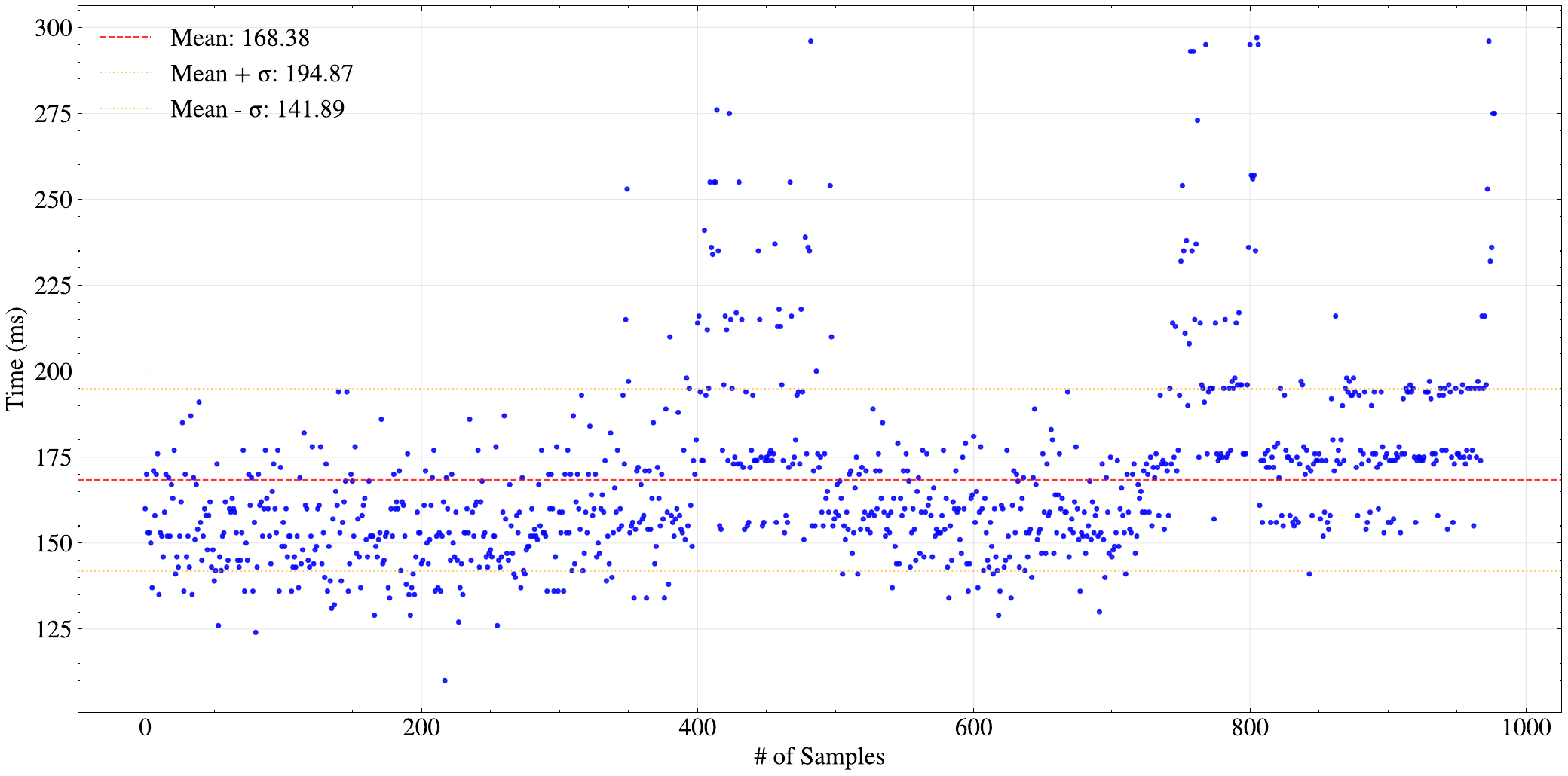}
    \caption{}
    \vspace{0pt}
    \label{fig:scatterg2g}
\end{subfigure}
\hfill
\begin{subfigure}[t]{.133\columnwidth}
    \centering
    \raisebox{7pt}{\includegraphics[width=\textwidth, height=0.55\textheight, keepaspectratio]{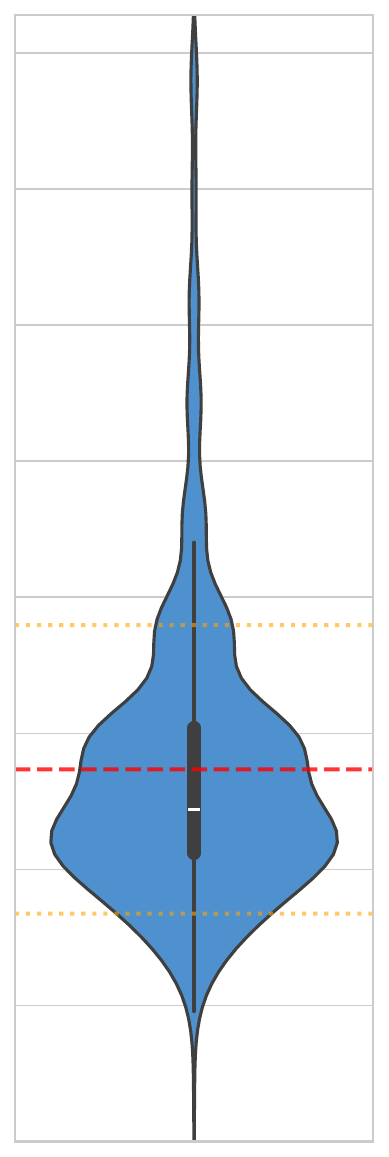}}
    \caption{}
    \label{fig:violing2g}
\end{subfigure}
    \caption{Glass-to-glass latency measures: sequence plot (a) and violin distribution (b).}
    \label{fig:plotsg2g}
\end{figure}

As shown in~\Cref{fig:plotsg2g}, the majority of the measured values are consistently concentrated between $150$ and $160$ ms. Occasional outliers with higher or lower latency values are observed, attributable to the limited precision and noise characteristics of the low-cost sensors employed in the experimental setup. These values, when added to the typical latencies of DSRC and 5G technologies (approximately $2-3$ ms and $18-20$ ms, respectively), ensure optimal latency in the remote driving scenario.

\section{Conclusion}

This paper presented an extensive experimental assessment of the feasibility of remote driving over heterogeneous wireless networks, focusing on latency, reliability, and coverage, which are critical for safe teleoperation.

The experimental results clearly show that ITS-G5 outperforms 5G in terms of latency whenever reliable radio coverage is available. Time-delta measurements indicate that DSRC packets arrive earlier than their cellular counterparts in more than 96\% of the cases, confirming the suitability of short-range V2I communications for latency-sensitive remote driving applications. In addition, DSRC exhibits low and stable packet-loss rates and maintains consistent uplink bitrate performance within RSU coverage areas. Conversely, cellular connectivity provides broader geographic coverage and more stable service continuity, albeit at the cost of higher, more variable latency, particularly in congested or transitional areas.

The coverage analysis further highlights the complementary nature of the two technologies. While DSRC delivers excellent performance in proximity to RSUs, its effectiveness is strongly influenced by line-of-sight conditions and infrastructure density. Cellular networks, on the other hand, ensure broader but more heterogeneous coverage due to external factors such as network load and urban signal attenuation. These findings strongly support the adoption of a hybrid communication strategy for remote driving, in which DSRC is used to minimize latency in critical zones, and cellular connectivity serves as a robust fallback to ensure uninterrupted service.

Glass-to-Glass latency consistently falls within the $150–160 ms$ range. When accounting for network overhead (approx. $2–18 ms$), the total end-to-end latency remains within the viable $162–178 ms$ window, aligning with functional thresholds for safe remote driving. These results confirm that the proposed architecture, integrated with edge processing and deployed over a live ITS network, meets the stringent temporal constraints of teleoperation; target performance is reliably achieved, particularly when DSRC serves as the primary communication link.

Overall, this work provides concrete experimental evidence that remote driving is feasible in real urban environments using existing ITS-G5 and cellular technologies, provided that a hybrid communication approach is adopted.
\balance
\bibliographystyle{IEEEtran}
\bibliography{references-short}

\end{document}